\documentclass{elsart}
\usepackage{amsfonts}
\usepackage{amsmath}
\usepackage{amssymb}
 
\usepackage{bm}
\usepackage{graphicx}

\begin{document}

\begin{frontmatter}

\title{Neutrino emission from spin waves in neutron spin-triplet superfluid}

\author{L. B. Leinson}
\address{Institute of Terrestrial Magnetism, Ionosphere and
 Radio Wave Propagation RAS, 142190 Troitsk, Moscow Region, Russia}

\begin{abstract}
The linear response of a neutron spin-triplet superfluid onto external weak axial-vector field is studied for the case of $^{3}P_{2}$ pairing with a projection of the total angular momentum $m_{j}=0$. The problem is considered in BCS approximation discarding Fermi-liquid effects. The anomalous axial-vector vertices of neutron quasiparticles possess singularities at some frequencies which specify existence of undamped spin-density waves in the Cooper condensate.  The spin waves are of a low excitation energy and are kinematically able to decay into neutrino pairs through neutral weak currents. The calculation predicts significant energy losses from within a neutron star at lowest temperatures when all other mechanisms of neutrino emission are killed by the neutron and proton superfluidity.  
\end{abstract}


\end{frontmatter}

When the temperature inside a neutron star core has dropped below the critical 
temperature $T_{c}$ for spin-triplet neutron pairing, the bulk baryon matter is 
expected to develop a superfluid condensate of neutrons \cite{Tamagaki}-\cite{Schwenk}, 
which has thermal excitations in the form of broken Cooper pairs. It is generally
accepted that neutrino emission from the pair-recombination processes in the
neutron triplet superfluid dominates the neutrino emissivities in many
cases \cite{YKL}. According to the minimal cooling paradigm \cite{Page04},
 \cite{Page09}, along with lowering of the temperature the star 
continues to lose its
energy by radiating low-energy neutrinos via the recombination processes
untill enters a photon-cooling epoch at $T\sim 0.1T_{c}$. At this latest stage of 
the neutrino-cooling, all other mechanisms of the neutrino emission from the inner 
core are suppressed greatly by the neutron and proton superfluidity \cite{YL}.  

In this paper we calculate the new neutrino emission process that can dominate 
in the temperature range typical for the final stage of neutrino cooling. 
Namely, we consider the weak decays of spin density excitations in
the superfluid triplet condensate of neutrons. 
Previously spin modes have been thoroughly studied in the p-wave superfluid liquid $^3He$ \cite{Maki}-\cite{Wolfle}. The pairing interaction in $^3He$  is invariant with respect to rotation of spin and orbital coordinates separately. This admits spin fluctuations independent of the orbital coordinates. In contrast, the spin-triplet neutron condensate arises in the high-density neutron matter due mostly to spin-orbit interactions which do not possess the above symmetry. Therefore the results obtained for liquid $^3He$ can not be applied directly to the spin-triplet neutron superfluid. 

 Recently spin waves with the excitation energy smaller than the superfluid energy gap was predicted to exist in the superfluid spin-triplet condensate of neutrons \cite{L09a}. 
The neutrino decay of such spin waves could be important for thermally-emitting neutron stars, presumably 
cooling through the combination of neutrino emission from the interior and photon 
cooling from the surface, the latter is responsible for their observed thermal 
emissions \cite{PZ02}. 

Let us remind briefly the theory of spin density
excitations in the condensate.
The triplet order parameter, $\hat{D}\equiv D_{\alpha \beta }\left( \mathbf{n%
}\right) $, in the neutron superfluid represents a symmetric matrix in spin
space $\left( \alpha ,\beta =\uparrow ,\downarrow \right) $ which can be
written as $\hat{D}\left( \mathbf{n}%
\right) =\Delta \mathbf{\bar{b}}\bm{\hat{\sigma}}\hat{g}$, where $%
\bm{\hat{\sigma}}=\left( \hat{\sigma}_{1},\hat{\sigma}_{2},\hat{\sigma}%
_{3}\right) $ are Pauli spin matrices, and $\hat{g}=i\hat{\sigma}_{2}$. The
gap amplitude $\Delta $ is a complex constant (on the Fermi surface), and $%
\mathbf{\bar{b}}\left( \mathbf{n}\right) $ is a real vector which we
normalize by the condition\footnote{In what follows we use the Standard Model of weak interactions, the
system of units $\hbar=c=1$ and the Boltzmann constant $k_{B}=1$.}
\begin{equation}
\int \frac{d\mathbf{n}}{4\pi }\bar{b}^{2}\left( \mathbf{n}\right) =1~.
\label{Norm}
\end{equation}%
The angular dependence of the order parameter is represented by the unit
vector $\mathbf{n=p}/p$ which defines the polar angles $\left( \theta
,\varphi \right) $ on the Fermi surface.

The neutron pairing at high densities involves a mixing of $^{3}P_{2}$ 
and $^{3}F_{2}$ channels \cite{Takatsuka}-\cite{Khodel}. Conventionally this 
mixing is not 
taken into account in estimates of neutrino emission from the neutron superfluid. 
Accordingly, throughout this paper we consider the case of $^{3}P_{2}$ neutron 
pairing, when quasiparticles pair in the most attractive channel with spin, 
orbital and total angular momenta, $ s=1,l=1,j=2$, respectively. 
Then the pairing interaction, in the most attractive channel, can be written as 
\cite{Tamagaki}
\begin{equation}
\nu \Gamma _{\alpha \beta ,\gamma \delta }\left( \mathbf{p,p}^{\prime
}\right) =-V\left( p,p^{\prime }\right) \sum_{m_{j}}\left( \mathbf{b}%
_{m_{j}}(\mathbf{n})\bm{\hat{\sigma}}\hat{g}\right) _{\alpha \beta }\left( 
\hat{g}\bm{\hat{\sigma}}\mathbf{b}_{m_{j}}^{\ast }(\mathbf{n}^{\prime
})\right) _{\gamma \delta }~,  \label{ppint}
\end{equation}%
where $V\left( p,p^{\prime }\right) $ is the corresponding
interaction amplitude; $\nu =p_{F}M^{\ast }/\pi ^{2}$ is the density of
states near the Fermi surface, and $\mathbf{b}_{m_{j}}\left( \mathbf{n}%
\right) $ are the vectors in spin space which generate standard spin-angle
matrices so, that 
\begin{equation}
\mathbf{b}_{m_{j}}(\mathbf{n})\bm{\hat{\sigma}}\hat{g}
\equiv\sum_{m_{s}+m_{l}=m_{j}}\left( \frac{1}{2}\frac{1}{2}\alpha\beta
|1m_{s}\right) \left( 11m_{s}m_{l}|2m_{j}\right) Y_{1,m_{l}}\left( \mathbf{n}%
\right)~.  \label{bm}
\end{equation}
These are given by 
\begin{eqnarray}
\mathbf{b}_{0} &=&\sqrt{1/2}\left( -n_{1},-n_{2},2n_{3}\right) ~,\mathbf{b}%
_{1}=-\sqrt{3/4}\left( n_{3},in_{3},n_{1}+in_{2}\right) ~,  \notag \\
\mathbf{b}_{2} &=&\sqrt{3/4}\left( n_{1}+in_{2},in_{1}-n_{2},0\right) ~,%
\mathbf{b}_{-m_{j}}=\left( -\right) ^{m_{j}}\mathbf{b}_{m_{j}}^{\ast }~
, \label{b012}
\end{eqnarray}%
where $n_{1}=\sin \theta \cos \varphi ,~n_{2}=\sin \theta \sin \varphi
,~n_{3}=\cos \theta $. The vectors are mutually orthogonal and are
normalized by the condition%
\begin{equation}
\int \frac{d\mathbf{n}}{4\pi }\mathbf{b}_{m_{j}^{\prime }}^{\ast }\mathbf{b}%
_{m_{j}}=\delta _{m_{j}m_{j}^{\prime }}.  \label{lmnorm}
\end{equation}%
We will focus on the condensation with $m_{j}=0$ which is conventionally
considered as the preferable one in the bulk matter of neutron stars \cite{Elg}, 
\cite{Page04}, \cite{Page09}. 
In this case $\mathbf{\bar{b}}\left( \mathbf{n}\right) =\mathbf{b}_{0}\left( 
\mathbf{n}\right) $.

Making use of the adopted graphical notation for the ordinary and anomalous propagators, $\hat{G}=\parbox{1cm}{\includegraphics[width=1cm]{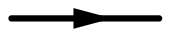}}$, $%
\hat{G}^{-}(p)=\parbox{1cm}{\includegraphics[width=1cm,angle=180]{Gn.eps}}$, 
$\hat{F}^{(1)}=\parbox{1cm}{\includegraphics[width=1cm]{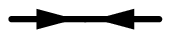}}$\thinspace
, and $\hat{F}^{(2)}=\parbox{1cm}{\includegraphics[width=1cm]{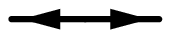}}$%
\thinspace , it is convenient to employ the Matsubara calculation technique for the system in thermal equilibrium.  Then the analytic form of the propagators is as
follows \cite{AGD}, \cite{Migdal}%
\begin{align}
\hat{G}\left( p_{m},\mathbf{p}\right) & =G\left( p_{m},\mathbf{p}\right)
\delta _{\alpha \beta }~,\ \ \ \ \ \ \ \hat{G}^{-}\left( p_{m},\mathbf{p}%
\right) =G^{-}\left( p_{m},\mathbf{p}\right) \delta _{\alpha \beta }~, 
\notag \\
\hat{F}^{\left( 1\right) }\left( p_{m},\mathbf{p}\right) & =F\left( p_{m},%
\mathbf{p}\right) \mathbf{\bar{b}}\bm{\hat{\sigma}}\hat{g}~,\ \ \ \hat{F}%
^{\left( 2\right) }\left( p_{m},\mathbf{p}\right) =F\left( p_{m},\mathbf{p}%
\right) \hat{g}\bm{\hat{\sigma}}\mathbf{\bar{b}}~,  \label{GF}
\end{align}%
where the scalar Green's functions are of the form $G^{-}\left( p_{m},%
\mathbf{p}\right) =G\left( -p_{m},-\mathbf{p}\right) $ and%
\begin{equation}
G\left( p_{m},\mathbf{p}\right) =\frac{-ip_{m}-\varepsilon _{\mathbf{p}}}{%
p_{m}^{2}+E_{\mathbf{p}}^{2}}~,\ F\left( p_{m},\mathbf{p}\right) =\frac{%
-\Delta }{p_{m}^{2}+E_{\mathbf{p}}^{2}}~.  \label{GFc}
\end{equation}%
Here $p_{m}\equiv i\pi \left( 2m+1\right) T$ with $m=0,\pm 1,\pm 2...$ is
the Matsubara's fermion frequency, and $\varepsilon _{\mathbf{p}%
}=p^{2}/\left( 2M^{\ast }\right) -p_{F}^{2}/\left( 2M^{\ast }\right) $ with $%
M^{\ast }=p_{F}/V_{F}$ being the effective mass of a quasiparticle. $%
\allowbreak \allowbreak $The quasiparticle energy is given by $E_{\mathbf{p}%
}^{2}=\varepsilon _{\mathbf{p}}^{2}+\Delta ^{2}\bar{b}^{2}\left( \mathbf{n}%
\right) $, where the (temperature-dependent) energy gap, $\Delta \bar{b}%
\left( \mathbf{n}\right) $, is anisotropic. In the absence of external
fields, the gap amplitude $\Delta $ is real.

Finally we introduce the following notation used below. We designate as $%
\mathcal{I}_{XX^{\prime }}\left( \omega ,\mathbf{n,q};T\right) $ the
analytical continuations onto the upper-half plane of complex variable $%
\omega $ of the following Matsubara sums:%
\begin{equation}
\mathcal{I}_{XX^{\prime }}\left( \omega _{n},\mathbf{n,q};T\right) \equiv
T\sum_{m}\frac{1}{2}\int_{-\infty }^{\infty }d\varepsilon _{\mathbf{p}%
}X\left( p_{m}+\omega _{n},\mathbf{p+}\frac{\mathbf{q}}{2}\right) X^{\prime
}\left( p_{m},\mathbf{p-}\frac{\mathbf{q}}{2}\right) ~.  \label{IXX}
\end{equation}%
where $X,X^{\prime }\in G,F,G^{-}$, and $\omega _{n}=2i\pi Tn$ with $n=0,\pm
1,\pm 2...$.These are functions of $\omega $, $\mathbf{q}$ and the direction
of the quasiparticle momentum $\mathbf{p}=p\mathbf{n}$.

We will focus on the
processes with $\omega ^{2}<2\Delta ^{2}\bar{b}^{2}$ and with a time-like
momentum transfer, $q^{2}<\omega ^{2}$. In this case the key role in the 
response theory belongs to the loop integral $\mathcal{I}_{FF}$.  
A straightforward
calculation yields $\mathcal{I}_{FF}\left( \mathbf{n},\omega ,\mathbf{qn}%
;T\right) =\mathcal{I}_{0}\left( \mathbf{n,}\omega ;T\right) +O\left(
q^{2}V_{F}^{2}/\omega ^{2}\right) $, where 
\begin{equation}
\mathcal{I}_{0}\left( \mathbf{n,}\omega ;T\right) =2\int_{0}^{\infty }\frac{%
d\varepsilon }{E}\frac{\Delta ^{2}}{4E^{2}-\left( \omega +i0\right) ^{2}}%
\tanh \frac{E}{2T}~,  \label{FFq0}
\end{equation}%
with 
\begin{equation*}
E=\sqrt{\varepsilon ^{2}+\Delta ^{2}\bar{b}^{2}}~\ .
\end{equation*}%
Insofar as $q^{2}V_{F}^{2}/\omega ^{2}\ll 1$ and $q^{2}V_{F}^{2}/\Delta
^{2}\ll 1$ we will neglect everywhere small corrections caused by a finite
value of space momentum $\mathbf{q}$.

We are interested in the linear medium response to the external axial-vector
field. The field interaction with a superfluid should be described with the
aid of four effective three-point vertices. There are two ordinary effective
vertices corresponding to creation of a particle and a hole by the field
(These differ by direction of fermion lines), and two anomalous vertices
corresponding to creation of two particles or two holes.

The anomalous effective vertices are given by the infinite sums of the
diagrams taking account of the pairing interaction in the ladder
approximation. In general, the ordinary vertices are to incorporate also  particle-hole interactions. Since Landau parameters for the
particle-hole interactions in asymmetric nuclear matter are unknown, we
simply neglect the Fermi-liquid effects and consider the effective
vertices and the pair correlation function in the BCS approximation. 
In this case the ordinary axial-vector vertices of
a particle and a hole are to be taken as $\bm{\hat{\sigma}}$
and $\bm{\hat{\sigma}}^{T}$, respectively.

\begin{figure}[h]
\includegraphics{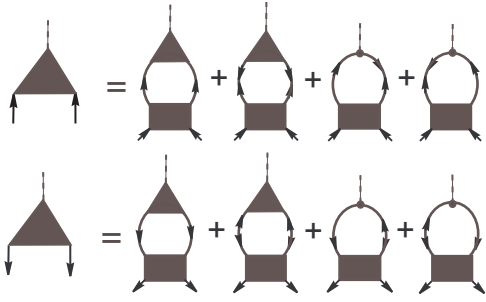}
\caption{Dyson's equations for the anomalous vertices. The shaded rectangle
represents the pairing interaction.}
\label{fig1}
\end{figure}

Given by the sum of ladder-type diagrams \cite{Larkin}, the anomalous
vertices are to satisfy the Dyson's equations symbolically depicted by
graphs in Fig. \ref{fig1}.

The vertex equations are to be  complemented by the gap equation which is 
of the following form \cite{Tamagaki} in the case $m_{j}=0$:
\begin{equation}
\Delta \left( p\right) =-\frac{1}{2\rho }\int dp^{\prime }p^{\prime
2}V\left( p,p^{\prime }\right) \Delta \left( p^{\prime }\right) \int \frac{d%
\mathbf{n}^{\prime }}{4\pi }{\frac{\bar{b}^{2}(\mathbf{n}^{\prime })}{2E(%
\mathbf{p}^{\prime })}}\tanh {\frac{E(\mathbf{p}^{\prime })}{2T}}~.
\label{GAP}
\end{equation}

We are interested in the processes occuring in a vicinity of the Fermi
surface. To get rid of the integration over the regions far from the Fermi surface we
renormalize the interaction as suggested in Ref. \cite{Leggett}: we define
\begin{equation}
V^{\left( r\right)}\left( p,p^{\prime};T\right)
=V\left( p,p^{\prime}\right) -V\left(
p,p^{\prime}\right) \left( GG^{-}\right) _{n}V^{\left(
r\right) }\left( p,p^{\prime};T\right) \ ,  \label{Vr}
\end{equation}
where the loop $\left( GG^{-}\right) _{n}$ is evaluated in the normal
(nonsuperfluid) state. In terms of $V^{\left( r\right) }$ the gap equation 
can be reduced to the following simple form
\begin{equation}
1=-V^{\left(
r\right)}\int \frac{d%
\mathbf{n}}{4\pi}
\bar{b}^{2}(\mathbf{n})A\left( \mathbf{n}\right) \ ,  \label{gapeq}
\end{equation}%
assuming that in the narrow vicinity of the Fermi surface the smooth
functions $V\left(p,p^{\prime }\right)$ and $\Delta\left( p^{\prime}\right)$ 
may be replaced with constants.

The function $A\left( \mathbf{n}\right)$ arises due to the renormalization 
procedure. It is given by
\begin{equation}
A\left( \mathbf{n}\right) =\frac{1}{2}\int_{0}^{\infty }d\varepsilon
\left( \frac{1}{\sqrt{\varepsilon ^{2}+\Delta ^{2}\bar{b}^{2}}}\tanh \frac{%
\sqrt{\varepsilon ^{2}+\Delta ^{2}\bar{b}^{2}}}{2T}-\frac{1}{\varepsilon }%
\tanh \frac{\varepsilon }{2T}\right) ~.  \label{An}
\end{equation}

Details of the analytic calculation can be found in Ref. \cite{L09a}, where 
after the proper renormalization the equations for the axial-vector anomalous 
vertices are obtained in the following analytic form:
\begin{align}
\hat{T}_{i}^{\left( 1\right) }\left( \mathbf{n}\right) & =V^{\left( r\right)}
\sum_{m_{j}}
\bm{\hat{\sigma}}\mathbf{b}_{m_{j}}(\mathbf{n})\hat{g}\int \frac{d\mathbf{n}%
^{\prime }}{8\pi }\left[ \mathcal{I}_{GG^{-}}\mathrm{Tr}\left( \hat{g}\left( %
\bm{\hat{\sigma}}\mathbf{b}_{m_{j}}^{\ast }\right) \hat{T}_{i}^{\left(
1\right) }\right) \right.  \notag \\
& -\mathcal{I}_{FF}\mathrm{Tr}\left( \left( \bm{\hat{\sigma}}\mathbf{b}%
_{m_{j}}^{\ast }\right) \left( \bm{\hat{\sigma}}\mathbf{\bar{b}}\right) \hat{%
g}\hat{T}_{i}^{\left( 2\right) }\left( \bm{\hat{\sigma}}\mathbf{\bar{b}}%
\right) \right)  \notag \\
& \left. -\frac{\omega }{\Delta }\mathcal{I}_{FF}2i\left( \mathbf{b}%
_{m_{j}}^{\ast }\mathbf{\times \bar{b}}\right) _{i}\right] _{\mathbf{n}%
^{\prime }}~,  \label{T1i}
\end{align}%
\begin{align}
\hat{T}_{i}^{\left( 2\right) }\left( \mathbf{n}\right) & =V^{\left( r\right)}
\sum_{m_{j}}\hat{g}\bm{\hat{\sigma}}\mathbf{b}_{m_{j}}^{\ast }(\mathbf{n})
\int \frac{d\mathbf{n%
}^{\prime }}{8\pi }\left[ \mathcal{I}_{G^{-}G}\mathrm{Tr}\left( \left( %
\bm{\hat{\sigma}}\mathbf{b}_{m_{j}}\right) \hat{g}\hat{T}_{i}^{\left(
2\right) }\right) \right.  \notag \\
& -\mathcal{I}_{FF}\mathrm{Tr}\left( \left( \bm{\hat{\sigma}}\mathbf{b}%
_{m_{j}}\right) \left( \bm{\hat{\sigma}}\mathbf{\bar{b}}\right) \hat{T}%
_{i}^{\left( 1\right) }\hat{g}\left( \bm{\hat{\sigma}}\mathbf{\bar{b}}%
\right) \right)  \notag \\
& \left. -\frac{\omega }{\Delta }\mathcal{I}_{FF}2i\left( \mathbf{b}_{m_{j}}%
\mathbf{\times \bar{b}}\right) _{i}\right] _{\mathbf{n}^{\prime }}~.
\label{T2i}
\end{align}%

The consistent solution to equations (\ref{gapeq}), (\ref{T1i}), 
and (\ref{T2i}) is found to be \cite{L09a}:
\begin{equation}
\mathbf{\hat{T}}^{\left( 1\right) }\left( \mathbf{n;}\omega ,T\right)
=f\left( \omega ,T\right) \left( \mathbf{e}\left( \bm{\hat{%
\sigma}b}_{1}\right) +\mathbf{e}^{\ast }\left( \bm{\hat{\sigma}b}%
_{-1}\right) \right) \hat{g}~,  \label{T1f}
\end{equation}%
\begin{equation}
\mathbf{\hat{T}}^{\left( 2\right) }\left( \mathbf{n;}\omega ,T\right)
=f\left( \omega ,T\right) \hat{g}\left( \mathbf{e}\left( 
\bm{\hat{\sigma}b}_{1}\right) +\mathbf{e}^{\ast }\left( \bm{\hat{%
\sigma}b}_{-1}\right) \right) ~,  \label{T2f}
\end{equation}%
where $\mathbf{e}=\left( 1,-i,0\right)$ is a constant complex vector in
spin space; and the function $f\left( \omega ,T\right) $ is given by the
expression
\begin{equation}
f\left( \omega ,T\right) =~\frac{1}{\chi \left( \omega
,T\right) }~\sqrt{\frac{3}{2}}\frac{\omega }{2\Delta }~\int \frac{d\mathbf{n}}{4\pi }%
\bar{b}^{2}\mathcal{I}_{FF}\left(\bar{b}^{2},\omega , T \right) ~
\label{fEq}
\end{equation}%
with%
\begin{eqnarray}
\chi \left( \omega ,T\right) &\equiv &\int \frac{d\mathbf{n}}{4\pi }~\allowbreak \left( \mathbf{b}_{1}^{\ast }\mathbf{b}_{1}-\bar{b}^{2}\right) A\left(\bar{b}^{2},\omega , T \right)  \notag \\
&& +2\int \frac{d\mathbf{n}}{4\pi }\left[ \frac{\omega ^{2}}{%
4\Delta ^{2}}\left( \mathbf{b}_{1}^{\ast }\mathbf{b}_{1}\right) -~\left( 
\mathbf{b}_{1}^{\ast }\mathbf{\bar{b}}\right) \left( \mathbf{\bar{b}b}%
_{1}\right) \right] \mathcal{I}_{FF}\left(\bar{b}^{2},\omega , T \right) ~.  \label{hi0}
\end{eqnarray}
In this expression the vectors $\mathbf{b}_{1}$, $\mathbf{b}_{1}^{\ast }$, and $\mathbf{\bar{b}}=\mathbf{b}_{0}$ are given by Eqs. (\ref{b012}).
As is well known, poles of the vertex function correspond to collective
eigen-modes of the system. Eigen frequencies, $\omega =\omega _{s}$, of such oscillations satisfy the equation $\chi \left(\omega _{s},T\right) =0$.
Since we consider
the time-like domain, $q<\omega _{s}$, the correction caused by finite value
of the wave number $\mathbf{q}$ is proportional to $q^{2}V_{F}^{2}\ll \omega
_{s}^{2}\left( 0\right) $. We will neglect this small (positive) correction
to the dispersion of spin waves. 

Thus the dispersion equation for the collective oscillations is of the form:
\begin{eqnarray}
&\int \frac{d\mathbf{n}}{4\pi }&\left( \mathbf{b}_{1}^{\ast }
\mathbf{b}_{1}-\bar{b}^{2}\right) A\left(\bar{b}^{2},\omega , T \right)  
\notag \\
&& +2\int \frac{d\mathbf{n}}{4\pi }\left[ \frac{\omega ^{2}}{%
4\Delta ^{2}}\left( \mathbf{b}_{1}^{\ast }\mathbf{b}_{1}\right) -~\left( 
\mathbf{b}_{1}^{\ast }\mathbf{\bar{b}}\right) \left( \mathbf{b}_{1}\mathbf{\bar{b}}\right) \right] \mathcal{I}_{FF}\left(\bar{b}^{2},\omega , T \right)=0 ~.  \label{deq}
\end{eqnarray}

Before proceeding to the detailed solution of Eq. (\ref{deq}), let us note that the equilibrium order parameter is specified in this equation solely by means of the real vector $\mathbf{\bar{b}}$. Therefore this equation allows to obtain the eigen energy of a similar excitation in a superfluid $^{3}He$-$B$ (Balian-Werthamer phase). 
In this case the order parameter matrix is given by $\hat{D}\left( \mathbf{n}\right) =\Delta \mathbf{\bar{b}}\bm{\hat{\sigma}}\hat{g}$ with $\mathbf{\bar{b}}=\mathbf{n}$, and the energy gap is isotropic \cite{BW}. The latter means that, in Eq. (\ref{deq}), the functions $A$ and $\mathcal{I}_{FF}$ are isotropic and can be moved beyond the integrals. Using also Eqs. (\ref{Norm}), (\ref{lmnorm}) we find
\begin{equation}
\int \frac{d\mathbf{n}}{4\pi }\left( \mathbf{b}_{1}^{\ast }
\mathbf{b}_{1}-\mathbf{\bar{b}}^{2}\right)=0 ~,  \label{A0}
\end{equation}
thus obtaining the dispersion equation
\begin{equation}
\int \frac{d\mathbf{n}}{4\pi }\left[ \frac{\omega ^{2}}{%
4\Delta ^{2}}\left( \mathbf{b}_{1}^{\ast }\mathbf{b}_{1}\right) -~\left( 
\mathbf{b}_{1}^{\ast }\mathbf{n}\right) \left( \mathbf{b}_{1}\mathbf{n}
\right) \right] =0 ~, \label{He}
\end{equation} 
which has the solution $\omega=\sqrt{8/5}\Delta$, in accordance with the energy of the spin wave in $^{3}He$-$B$ obtained in \cite{Wolfle}.

We now return to examination of spin waves in the spin-triplet neutron superfluid. A simple estimate of the excitation energy of the wave can be made using the angle-averaged energy gap in the quasiparticle energy, $\left\langle \Delta ^{2}\bar{b}^{2}\right\rangle \equiv \Delta ^{2}$. In this approximation the functions  $A$ and $\mathcal{I}_{FF}$ are isotropic and can be moved beyond the integrals. Using also Eqs. (\ref{A0}) we obtain the equation
\begin{equation}
\int \frac{d\mathbf{n}}{4\pi }\left[ \frac{\omega ^{2}}{%
4\Delta ^{2}}\left( \mathbf{b}_{1}^{\ast }\mathbf{b}_{1}\right) -~\left( 
\mathbf{b}_{1}^{\ast }\mathbf{\bar{b}}\right) \left( \mathbf{b}_{1}\mathbf{\bar{b}}\right) \right] =0 ~, \label{Nm}
\end{equation} 
with $\mathbf{\bar{b}}=\mathbf{b}_0$. This equation has the solution $\omega=\Delta / \sqrt{5}$. This simple estimate shows that the energy of the spin wave excitation is smaller than the energy gap in the quasiparticle spectrum. Neutrino decays of such spin waves could be important for thermally-emitting neutron stars. Therefore let us  examine the wave excitation energy more accurately.

Since the functions $A\left(\bar{b}^{2},\omega , T \right)$ and $\mathcal{I}_{FF}\left(\bar{b}^{2},\omega , T \right)$ are axially symmetric one can integrate Eq. (\ref{deq}) over azimuth angle. We then obtain the equation
\begin{eqnarray}
&\int_{0}^{1}& dn_{3}\left(
1-3n_{3}^{2}\right) A\left( n_{3}\right)  \notag \\
&&=-3\int_{0}^{1}dn_{3}\left( \frac{\omega ^{2}}{2\Delta ^{2}}
\left( 1+n_{3}^{2}\right) -~n_{3}^{2}\left( 1-n_{3}^{2}\right) \right) 
\mathcal{I}_{0}\left( n_{3},\omega ,T\right) ~.  \label{hi}
\end{eqnarray}%
It is convenient to define the dimensionless variables
$y\equiv \Delta \left( T\right) /T$
and $\Omega =\omega /\left( 2\Delta \right) $.
We can get a clear idea of the behavior of the left side of Eq. (\ref{hi})
as a function of $\Omega $ near the critical temperature.
Obviously, $y$ tends to zero as the temperature nears the critical value, 
$T\rightarrow T_{c}$. In this case we find $A\left( n_{3};T\right)
\rightarrow -\varkappa y^{2}b^{2}$, where%
\begin{equation*}
\varkappa \equiv \frac{1}{16}\int_{0}^{\infty }\allowbreak \frac{dx}{x^{3}}%
\left( \tanh x-\allowbreak x\left( 1-\tanh ^{2}x\allowbreak \right)
\allowbreak \right) =5.32\times 10^{-2},
\end{equation*}%
In the same limit $T\rightarrow T_{c}$, the function $\mathcal{I}_{0}$, tends to 
\begin{equation*}
\mathcal{I}_{0}\left( n_{3},\Omega ;y\right) \rightarrow \frac{\pi }{8}\frac{%
y}{\sqrt{\bar{b}^{2}-\Omega ^{2}}}
\end{equation*}%
We obtain the dispersion equation of the form 
\begin{equation*}
\allowbreak \Omega _{s}^{2}\int_{0}^{1}dn_{3}\frac{2\left( 1+n_{3}^{2}\right)}{%
\sqrt{\left( 1+3n_{3}\right) ^{2}-2\Omega _{s}^{2}}}-~\int_{0}^{1}dn_{3}%
\frac{n_{3}^{2}\left( 1-n_{3}^{2}\right) }{\sqrt{\left( 1+3n_{3}\right)
^{2}-2\Omega _{s}^{2}}}=-0.02y
\end{equation*}

According to the above estimate $\Omega _{s}^{2}\ll 1$. Therefore the equation can be solved by iterations. To the lowest accuracy we find 
\begin{equation*}
\Omega _{s}^{2}~\simeq 4.28\times 10^{-2}-1.76\times 10^{-2}y
\end{equation*}%

\begin{figure}[h]
\includegraphics{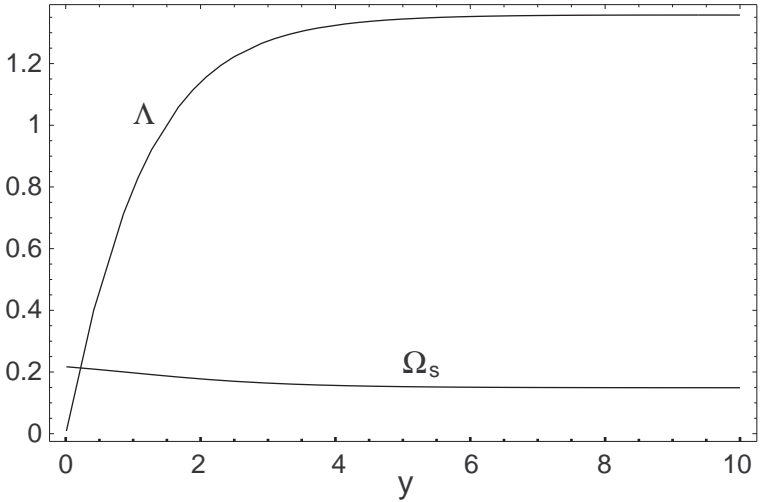}
\caption{Dimensionless frequency $\Omega _{s}=\protect\omega _{s}\left(
0\right) /2\Delta $ and function $\Lambda $ vs. the temperature parameter $%
y=\Delta \left( T\right) /T$.}
\label{fig3}
\end{figure}

For arbitrary $y>0$ the dispersion equation requires numerical computations. 
In Fig. \ref{fig3}, the dimensionless frequency 
$\Omega _{s}=\omega _{s}\left( 0\right) /\left( 2\Delta \right) $ of the 
collective spin oscillations is shown versus the dimensionless parameter 
$y\equiv \Delta \left( T\right) /T$.

Just below the critical temperature, $y\rightarrow 0$, we obtain $\Omega
_{s}\simeq 0.21$, or, identically, $\omega _{s}\left(
T\rightarrow T_{c}\right) \simeq 0.42\Delta $. The spin wave frequency
slightly decreases along with lowering of the temperature and tends to a
plateau at $\Omega _{s}\simeq \allowbreak 0.15$, what is equivalent to $%
\omega _{s}\left( T\rightarrow 0\right) \simeq 0.3\Delta $.

\begin{figure}[h]
\includegraphics{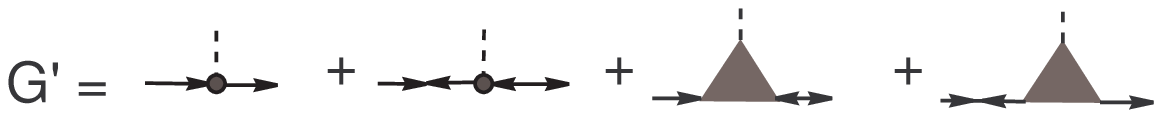}
\caption{Correction to the ordinary propagator of a quasiparticle in
external field.}
\label{fig4}
\end{figure}

Since the spin excitations occur in the time-like domain of the transferred
energy and momentum the decay into neutrino-pairs is kinematically allowed.
We examine the neutrino energy losses in the standard model of weak
interactions. Then after integration over the phase volume of freely
escaping neutrinos and antineutrinos the total energy which is emitted per
unit volume and time can be obtained in the form (see details, e.g., in Ref. 
\cite{L01}):
\begin{equation}
\epsilon =-\frac{G_{F}^{2}\mathcal{N}_{\nu }}{192\pi ^{5}}\int_{0}^{\infty
}d\omega \int d^{3}q\frac{\omega \Theta \left( \omega -q\right) }{\exp
\left( \omega \slash {T}\right) -1}\operatorname{Im}\Pi _{\mathrm{weak}}^{\mu \nu
}\left( \omega ,\mathbf{q}\right) \left( k_{\mu }k_{\nu }-k^{2}g_{\mu \nu
}\right) ~,  \label{QQQ}
\end{equation}%
where $G_{F}$ is the Fermi coupling constant, $\mathcal{N}_{\nu }=3$ is the
number of neutrino flavors, $k^{\mu }\equiv \left( \omega ,\mathbf{q}\right) 
$, $\Theta \left( x\right) $ is Heaviside step function, 
and $\Pi _{\mathrm{weak}}^{\mu \nu }$ is the retarded weak polarization
tensor of the medium. The latter can be found \cite{Migdal} using the linear 
correction to the Green function of a quasiparticle 
$\mathbf{\hat{G}}^{\prime }=\delta\hat{G} \slash \delta\mathbf{A}$  caused by 
the external field $\mathbf{A}$. In the BCS approximation, it is given by the diagrams shown in
Fig. \ref{fig4}, 
and can be written analytically as 
\begin{equation}
\mathbf{\hat{G}}^{\prime }=GG~\bm{\hat{\sigma}}+FF~\left( \bm{\hat{\sigma}}%
\mathbf{\bar{b}}\right) \bm{\hat{\sigma}}\left( \bm{\hat{\sigma}}\mathbf{%
\bar{b}}\right) +GF~\mathbf{\hat{T}}^{\left( 1\right) }\hat{g}\left( %
\bm{\hat{\sigma}}\mathbf{\bar{b}}\right) +FG~\left( \bm{\hat{\sigma}}\mathbf{%
\bar{b}}\right) \hat{g}\mathbf{\hat{T}}^{\left( 2\right) }~,  \label{Gpr}
\end{equation}%
where $GG\equiv G\left( p_{m}+\omega _{n},\mathbf{p+q}/2\right) G\left(
p_{m},\mathbf{p-q}/2\right) $, ect.

In the axial channel, the pair correlation function can be found as the
analytic continuation of the following Matsubara sum%
\begin{equation}
\Pi _{\mathrm{A}}^{ij}\left( \omega _{n},q\right) =T\sum_{m}\int \frac{d^{3}%
\mathbf{p}}{8\pi ^{3}}\mathrm{Tr}\left( \sigma _{i}\hat{G}_{j}^{\prime
}\right) ~,  \label{K}
\end{equation}%
and the imaginary part of the weak polarization tensor can be written as $%
\operatorname{Im}\Pi _{\mathrm{weak}}^{\mu \nu }=\delta ^{\mu i}\delta ^{\nu
j}C_{A}^{2}\operatorname{Im}\Pi _{\mathrm{A}}^{ij}$, where $C_{A}\simeq 1.26$
is the axial-vector weak coupling constant for neutrons.

To the lowest accuracy in $V_{F}\ll 1$, we may evaluate the polarization tensor
in the limit $\mathbf{q}=0$. Then using Eqs. (\ref{T1f}), (\ref{T2f}), (\ref%
{Gpr}), and (\ref{K}) we find after some algebraic manipulations (Dependence on $\mathbf{n}$ and $\omega $ in the integrand is omitted for brevity.):%
\begin{align}
\Pi _{\mathrm{A}}^{ij}\left( \omega \right) & =4\rho \int \frac{d\mathbf{n}}{%
4\pi }\left[ \frac{1}{2}\left( \mathcal{I}_{GG}-\bar{b}^{2}\mathcal{I}%
_{FF}\right) \delta _{ij}+\bar{b}^{2}\mathcal{I}_{FF}\frac{\bar{b}_{i}\bar{b}%
_{j}}{\bar{b}^{2}}\right.  \notag \\
& \left. +\left( \delta _{ij}-\delta _{i3}\delta _{j3}\right) \frac{3}{4}%
\frac{\sqrt{6}}{2}\frac{\omega }{2\Delta }\ f\left( \omega \right) \bar{b}%
^{2}\mathcal{I}_{FF}\right] _{\mathbf{q}=0}~.  \label{Ka}
\end{align}%

Using the identity $\left( \mathcal{I}_{GG}-\bar{b}^{2}\mathcal{I}%
_{FF}\right) _{q\rightarrow 0}=-2\bar{b}^{2}\mathcal{I}_{0}$, which can be
verified by a straightforward calculation, we find 
\begin{equation}
\Pi _{\mathrm{A}}^{ij}=-4\rho \int \frac{d\mathbf{n}}{4\pi }\left( \delta
_{ij}-\frac{\bar{b}_{i}\bar{b}_{j}}{\bar{b}^{2}}-\frac{3}{4}\left( \delta
_{ij}-\delta _{i3}\delta _{j3}\right) \frac{\sqrt{6}}{2}\frac{\omega }{%
2\Delta }\ f\left( \omega \right) \right) \bar{b}^{2}\mathcal{I}_{0}\left(
\omega \right) ~.  \label{AKij}
\end{equation}

Imaginary part of this function consists of two contributions. The first one
arises from the imaginary part of $\mathcal{I}_{0}\left( \omega \right) $
and exists at $\omega >2\bar{b}\Delta $. This part is responsible for the 
emission of neutrino pairs from breaking and formation of the Cooper pairs 
at thermal equilibrium and is already discussed in Ref. \cite{L09a}. 
We are interested in the neutrino emission from the decay of spin waves. 
The corresponding contribution into the imaginary part of the axial polarization
tensor (\ref{AKij}) arises from the pole part of the function $f\left(
\omega \right) $ at $\chi \left( \omega \right) =0$, and can be found with
the aid of Sokhotsky's formula, $\left( \chi +i0\right) ^{-1}=\mathcal{P}%
\left( 1/\chi \right) -i\pi \delta \left( \chi \right) $. Taking into
account Eq. (\ref{deq}) we obtain: 
\begin{equation}
\operatorname{Im}f\left( \omega \right) =-\pi \sqrt{6}\frac{\omega }{2\Delta }\left(
\int_{0}^{1}dn_{3}\left( 1+3n_{3}^{2}\right) \mathcal{I}_{0}\right) \frac{1}{%
\left\vert \partial \chi /\partial \omega ^{2}\right\vert _{\omega _{s}}}%
\delta \left( \omega ^{2}-\omega _{s}^{2}\right)  \label{imf}
\end{equation}%
We remind that $\mathcal{I}_{0}$ is real function in the domain $0<\omega
<2\Delta \bar{b}$, where the undamped spin waves exist. To simplify Eq. (\ref%
{imf}) we use the fact that $\omega _{s}^{2}\ll 4E^{2}$. This allows to
neglect $\omega ^{2}$ in Eq. (\ref{FFq0}), and substitute 
\begin{equation*}
\mathcal{I}_{0}\simeq \frac{1}{2\bar{b}^{2}}\int_{0}^{\infty }\frac{du}{%
\left( u^{2}+1\right) ^{3/2}}\tanh \frac{y\bar{b}}{2}\sqrt{u^{2}+1}
\end{equation*}%
in Eqs. (\ref{AKij}) and (\ref{imf}). In obtaining this expression we used
the change $\varepsilon \equiv u\Delta \bar{b}$.

Then we find
\begin{equation*}
\operatorname{Im}f\left( \omega \right) =-\pi \frac{\Delta }{\sqrt{6}}\frac{%
\int_{0}^{1}dn_{3}\left( 1+3n_{3}^{2}\right) \mathcal{I}_{0}}{%
\int_{0}^{1}dn_{3}\allowbreak \left( 1+n_{3}^{2}\right) \mathcal{I}_{0}}%
\delta \left( \omega -\omega _{s}\right) ~,
\end{equation*}%
and the imaginary part of the weak polarization tensor can be written as 
\begin{equation}
\operatorname{Im}\Pi _{\mathrm{weak}}^{\mu \nu }
= -\delta ^{\mu i}\delta ^{\nu j}\left( \delta _{ij}-\delta _{i3}\delta
_{j3}\right) \frac{3\pi }{8}C_{A}^{2}\rho \omega \ \delta \left( \omega
-\omega _{s}\right) \Lambda \left( y\right) ~,  \label{ImP}
\end{equation}%
where 
\begin{equation*}
\Lambda \left( y\right) \equiv \frac{\left( \int_{0}^{1}dn_{3}\left(
1+3n_{3}^{2}\right) \mathcal{I}_{0}\right) ^{2}}{\int_{0}^{1}dn_{3}%
\allowbreak \left( 1+n_{3}^{2}\right) \mathcal{I}_{0}}~.
\end{equation*}%
As shown in Fig. 2, this function increases smoothly along with lowering of the
temperature and tends to plateau at $\Lambda =1.35$

Inserting Eq. (\ref{ImP}) into Eq. (\ref{QQQ}) and performing trivial
integrations we find the neutrino emissivity due to spin wave decays (SWD):
\begin{equation}
\epsilon _{SWD}=\frac{1}{5\pi ^{5}}G_{F}^{2}C_{A}^{2}\mathcal{N}_{\nu
}p_{F}M^{\ast }T^{7}F\left( y\right) ~,  \label{eps}
\end{equation}%
Here $y=\Delta \left( T\right) /T$, and%
\begin{equation*}
F\left( y\right) \equiv \frac{y^{7}\Lambda \left( y\right) \Omega
_{s}^{7}\left( y\right) }{\exp \left( 2y\Omega _{s}\left( y\right) \right) -1%
}~.
\end{equation*}%
For a practical usage we reduce Eq. (\ref{eps}) to the traditional form%
\begin{equation}
\epsilon _{SWD}\simeq 6.\,\allowbreak 3\times 10^{20}\frac{erg}{cm^{3}s}%
~\left( \frac{M^{\ast }}{M}\right) \left( \frac{p_{F}}{Mc}\right)
T_{9}^{7}C_{\mathrm{A}}^{2}F\left( \Delta \left( T\right) /T\right) ~,
\label{erg}
\end{equation}%
where $M$ and $M^{\ast }$ are the effective and bare nucleon masses,
respectively; $c$ is speed of light, and $T_{9}=T/\left( 10^{9}K\right) $.

The function $F\left( y\right) $ depends actually on the only parameter $y$
and can be easily evaluated making use of the analytic fits which relate $%
\Omega _{s}$ and $\Lambda $ to $y$ at any $y>0$: 
\begin{equation*}
\Omega _{s}\left( y\right) =\frac{0.2172-0.0059y+0.0114y^{2}+0.0026y^{3}}{%
1+0.0534y+0.0710y^{2}+0.0175y^{3}}~,
\end{equation*}%
\begin{equation*}
\Lambda \left( y\right) =\frac{1.1038y+0.2589y^{2}+0.1063y^{3}}{%
1+0.5653y+0.2094y^{2}+0.0780y^{3}}~.
\end{equation*}%
The maximum fit error is about 0.$1$\% both for $\Omega _{s}\left( y\right) $
and for $\Lambda \left( y\right) $.

It is convenient to fit also the function $y=\Delta \left( T\right) /T$ for
practical computations. For this purpose we can adjust the simple expression
suggested in Ref. \cite{YKL}, where 
\begin{equation}
\mathtt{\upsilon }\left( \tau \right) \mathtt{\equiv }\frac{\Delta
_{0}\left( T\right) }{T}=\sqrt{1-\tau }\left( 0.7893+\frac{1.188}{\tau }%
\right) ,  \label{fit}
\end{equation}%
and $\tau \equiv T/T_{c}$. In Ref. \cite{YKL}, the gap amplitude $\Delta
_{0}\left( T\right) $ is defined by the relation $\Delta _{\mathbf{n}%
}^{2}=\Delta _{0}^{2}\left( 1+3\cos ^{2}\theta \right) $, while our
definition is $\Delta _{\mathbf{n}}^{2}=\frac{1}{2}\Delta ^{2}\left( 1+3\cos
^{2}\theta \right) $, i.e. the gap amplitude $\Delta \left( T\right) $ is $%
\sqrt{2}$ times larger than the gap amplitude $\Delta _{0}\left( T\right) $
used in Ref. \cite{YKL}, and thus $y\left( \tau \right) =\sqrt{2}\mathtt{%
\upsilon }\left( \tau \right) $.

\begin{figure}[h]
\includegraphics{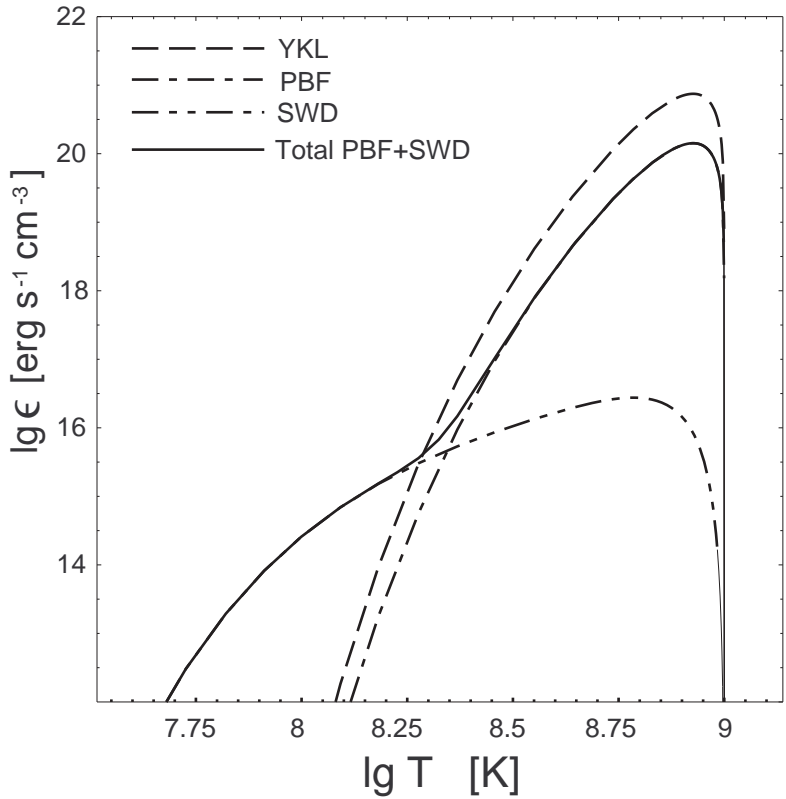}
\caption{Temperature dependence of the neutrino emissivity due to Cooper
pairing of neutrons and due to decay of spin waves at $\protect\rho =2\times
10^{14}~g~cm^{-3}$ and $T_{c}=10^{9}K$. For a comparison, we show the
emissivity as obtained in Ref. \protect\cite{YKL} (dash line). }
\label{fig5}
\end{figure}

The result of numerical computation is shown in Fig. \ref{fig5}, where we
compare the neutrino emissivity due to spin wave decays with the
self-consistent emissivity from the pair breaking and formation processes
(PBF), as calculated in Ref. \cite{L09a}. For a comparison we demonstrate
the PBF energy losses ignoring the anomalous weak interactions, as suggested
in Ref. \cite{YKL}. The temperature dependence of the emissivity is
evaluated at $\rho =2\times 10^{14}~g~cm^{-3}$. We set the effective nucleon
masses $M^{\ast }=0.7M$, for simplicity. The critical temperature for
neutron pairing is chosen to be $T_{c}=10^{9}K$. Total of the energy losses, 
$\epsilon =\epsilon _{\mathrm{PBF}}+\epsilon _{\mathrm{SWD}}$, is also
shown. One can see that the decay of spin waves into neutrino pairs is very
effective at low temperatures, when other known mechanisms of neutrino
energy losses in the bulk neutron matter are strongly suppressed by
superfluidity. 

Let us summarize our results. We have studied the response functions of the
triplet neutron superfluid onto external axial-vector field with taking into
account of anomalous interactions. The calculations are made in the BCS
approximation for the case of $^{3}P_{2}$ superfluid condensate with $%
m_{j}=0 $.

Our theoretical analysis predicts the existence of spin-density waves of a
small excitation energy 
$\omega _{s}\simeq \left( 0.3\div 0.4\right)\Delta $ in superfluid
condensate. We have calculated the neutrino energy losses caused by the
decay of the spin waves through neutral weak currents. Neutrino energy
losses due to spin wave decays are given by Eq. (\ref{eps}). Because of a
rather small excitation energy the decay of spin waves leads to a
significant neutrino emission at lowest temperatures $T\ll T_{c}$, when all
other mechanisms of the neutrino energy losses are killed by a
superfluidity. Since the neutron $^{3}P_{2}$ pairing occurs in the core
which contains more than 90 percents of the neutron star volume, the decay
of spin waves can affect the minimal cooling scenario. In the enhanced
cooling scenario, the effect of spin wave decays could be also quite
noticeable if the proton (spin-singlet) superfluid suppresses greatly the
direct Urca process.

\end{document}